# Cryogenic phonon-scintillation detectors with PMT readout for rare event search experiments

X. Zhang[1], J. Lin[1], V. B. Mikhailik[2], and H. Kraus[1,*]
[1]*Department of Physics, University of Oxford, Keble Rd., Oxford, OX1 3RH, UK*
[2]*Diamond Light Source, Harwell Science Campus, Didcot, OX11 0DE, UK*

## Abstract

Cryogenic phonon-scintillation detectors (CPSD) for rare event search experiments require reliable, efficient and robust photon detectors that can resolve individual photons in a scintillation event. We report on a cryogenic detector containing a scintillating crystal, equipped with an NTD-Ge phonon sensor and a photon detector based on a low-temperature photomultiplier tube (PMT) that is powered by a Cockcroft–Walton generator. Here we present results from the characterisation of two detector modules, one with $CaWO_4$, the other with $CaMoO_4$ as scintillator. The energy resolutions (FWHM) at 122.1 keV for the scintillation / PMT channel are 19.9% and 29.7% respectively for $CaWO_4$ and $CaMoO_4$ while the energy resolutions (FWHM) for the phonon channels are 2.17 keV (1.8%) and 0.97 keV (0.79%). These characteristics compare favourably with other CPSDs currently used in cryogenic rare-event search experiments. The detection module with PMT readout benefits from the implementation of a well-understood, reliable, and commercially available component and improved time resolution, while retaining the major advantages of conventional CPSD, such as high sensitivity, resolving power and discrimination ability.

*- corresponding author

*Paper published in Astroparticle Physics, 79 (2016) 31*-40

## 1. Introduction

Within modern particle physics, rare event search experiments offer opportunities for discovering new physics complementary to high-energy physics experiments. The discovery of weakly interacting massive particles (WIMP) or neutrinoless double beta decay (DBD) could enhance significantly our understanding of the Universe. This motivates significant research effort, using different experimental techniques (see e.g. [1], [2] and references therein), all focussed on searches of such events. The fundamental limitations of any of these endeavours are defined by both the precision with which signals can be measured and the capability of distinguishing extremely rare signal events from various and copious backgrounds, typically caused by environmental and cosmogenic radioactivity.

Cryogenic phonon detectors, operating at temperatures well below 100 mK made possible the very precise determination of the energy deposited by a particle or ionising radiation. Currently these detectors can reach a resolution of better than 2 eV at 5 keV, far superior to any of the mainstream ionization detectors [3]. Further, cryogenic detectors can be equipped with powerful event type recognition through simultaneously recording the energy deposited as phonons and as scintillation [4], [5]. This is applied successfully in a number of experiments [6], [7]. The potential of this technique is demonstrated through the detection of the alpha-decay of long-lived isotopes $^{209}$Bi [8], $^{180}$W [9] and $^{151}$Eu [10], and by results from cryogenic experiments searching for WIMPs [11], [12], [13]. Moreover, it has been demonstrated that such hybrid cryogenic phonon-scintillation detectors (CPSD) are well suited to reach the sensitivity level required by future experimental quests for neutrinoless DBD [14], [15], [16], [17]. The greatest advantage of CPSD - the active background rejection - relies on the availability of efficient methods for detection of photons emitted by scintillators at cryogenic temperatures.

## 2. Detection of light at cryogenic temperatures

Due to fundamental losses, the main fraction of energy absorbed by a scintillator is converted into heat and only a small part (<10%, in general) is emitted at low temperatures as photons [18]. The heat channel is used for determining the energy deposited in the interaction while the response generated in the light channel determines the discrimination power of the CPSD [19]. High efficiency of photon detection is always a priority for a scintillation detector but achieving this goal at cryogenic temperatures is a very challenging task. Improving the effectiveness of scintillation detection can be achieved through maximising light collection efficiency, which is possible when the detector is in close proximity to the crystal scintillator. Given that the crystal is cooled to well below 100 mK for operating as low-temperature calorimeter, the only solution adopted so far was to use another phonon detector optimised for detection of low-energy visible photons [20]. Development of such light detectors for the needs of rare event search experiments has been a focus of research efforts over the last decade.

Superconducting transition-edge sensors (TES) can, in principle, be made sensitive to a few scintillation photons [21]. To reach optimal sensitivity, TES require precise stabilization of their operating temperature within the phase transition from the superconducting to normal state. The signal from a TES is read out using a superconducting quantum interference device (SQUID). Good examples of detectors developed for a cryogenic dark matter search experiment demonstrate an energy threshold of 3 keV [22] (though typically higher values were used as data analysis threshold [23]) and exhibit an energy resolution of 180 eV at 5.89 keV, the energy of the $K_\alpha$-photons of $^{55}$Mn [24]. It has to be noted that reliable production of sensors with the same properties of phase transitions is a very challenging task. Consequently, despite the excellent performance of such detectors, due to the lack of reproducibility and the specialist knowledge required to produce them, their wider-spread use remains limited.

Cryogenic light detectors with neutron transmutation doped (NTD) Ge-thermistors tend to exhibit higher energy thresholds [20], [25] but they can operate over a wider temperature range and their front-end readout can be realized with fairly cost-effective electronics. These detectors are well suited for the measurement of a scintillation response caused by high-energy particles and therefore they are ideal in searches for neutrinoless DBD [16]. An example of the energy resolution achieved for such a light detector in the latter application is 250 eV at 5.9 keV [26]. Research aiming to improve performance characteristics of TES and NTD detectors using Neganov-Luke amplification are underway [27], [28]. Recently, the development of a cryogenic light detector with a magnetic metal calorimeter (MMC) for neutrinoless DBD experiments has been reported [29]. The detector exhibits an energy resolution of 545eV for the $K_{\alpha}$-line of $^{55}$Mn. There is further an initiative aiming to develop cryogenic phonon-mediated kinetic inductance detectors with a large area as a light sensor [30].

The main advantage of a cryogenic detector is its sensitivity to the tiny changes of temperature caused by the absorption of photons or particles. However, as the detector is an integrating device, it is unavoidably very responsive to many different kinds of energy deposition. In particular it is very susceptible to mechanical vibrations that can cause excessive background noise [26]. For the light detector this noise is what determines the detector threshold and eventually influences the overall sensitivity of the experiment. Another disadvantage is the relatively large time constant of such a light detector (~1−10 ms) [22], [24], [26], leading to poor timing resolution of CPSD.

Implementation of single-photon light detectors can radically improve this situation. The need to have available the option of using a photon counting device capable of meeting the challenging requirements of cryogenic applications motivated extensive studies of different photodetectors down to temperatures of a few Kelvin. It has been shown that the performance of an avalanche photodiode [31] or a silicon photomultiplier [32] significantly degrades due to freeze-out of charge carriers. It is worth remarking that a silicon photomultiplier has been recently characterised down to the temperature of liquid He [33] but nonetheless the technology as it stands is suboptimal and the performance of the device as detector of scintillation at cryogenic temperatures still needs to be demonstrated.

In this context photomultiplier tubes (PMT) may potentially offer the simplest and commercially available solution. A PMT is a technologically mature product developed and optimised for detecting photons. PMTs remain the preferred option for the detection of scintillation due to the following reasons: i) excellent sensitivity for single photon detection (20-25%), ii) fast (~10 ns) responsivity iii) large sensitive area, iv) ruggedness and v) reliability of operation. That is why in the last decade major research efforts have been directed towards improving the performance of PMTs and widening the temperature range of their operation. The suitability of a modified (low-temperature) version of PMTs for operation at temperatures of liquid noble gases has already been confirmed by numerous studies [34], [35], [36] and PMTs are widely used in the relevant experiments (see e.g. [37] and reference therein). Recently it has been shown that such a PMT can reliably detect single photons when cooled close to the temperature of liquid He [38], [39], [40].

An important advantage of operating a detector in the photon counting regime is the much improved signal-to-noise ratio, especially in the low-frequency region, compared with an integrating light detector. At low temperature, the PMT noise (dark events) caused by spontaneous emission and afterpulses [41] is much less severe and, moreover, its impact can be reduced significantly if true events can be tagged for instance by an independent trigger. Another advantage of PMT readout in the case of measurements of scintillations is attributed to the fact that the timing characteristics of the detected signal is almost exclusively determined by the decay time characteristics of the light emission process in the scintillator (typically <1 ms). This allows improving the time resolution of the detector and consequently reducing the severity of pile-up - a potentially serious problem for relatively slow cryogenic detectors.

Altogether this provides motivation for trying PMT as a light detector for CPSD. In this study we successfully tested two CPSD modules (see Fig. 1). One was based on a $CaWO_4$ scintillator crystal as absorber, the other used $CaMoO_4$. Both calorimeters were equipped with an NTD-Ge phonon sensor and scintillation was detected with a low-temperature PMT from Hamamatsu (model R8520-06).

## 3. Experimental apparatus

The detector module is composed of two separate detectors (Fig.1), a low-temperature calorimeter and a PMT, to simultaneously measure phonon and scintillation signals, respectively. The detectors are mounted within a copper housing, lined with VM2000 (manufactured by 3M) highly reflecting foil **[42]**. The detector housing serves two main purposes: being a heat bath, which allows the calorimeter to return to the operating temperature after a particle interaction, and being the inner shielding against natural radioactive background. The scintillating absorber, either $CaWO_4$ ($3.3 \times 8.4 \times 9.7$ mm$^3$) or $CaMoO_4$ ($5.0 \times 9.5 \times 9.5$ mm$^3$), was suspended within the housing using nylon wires.

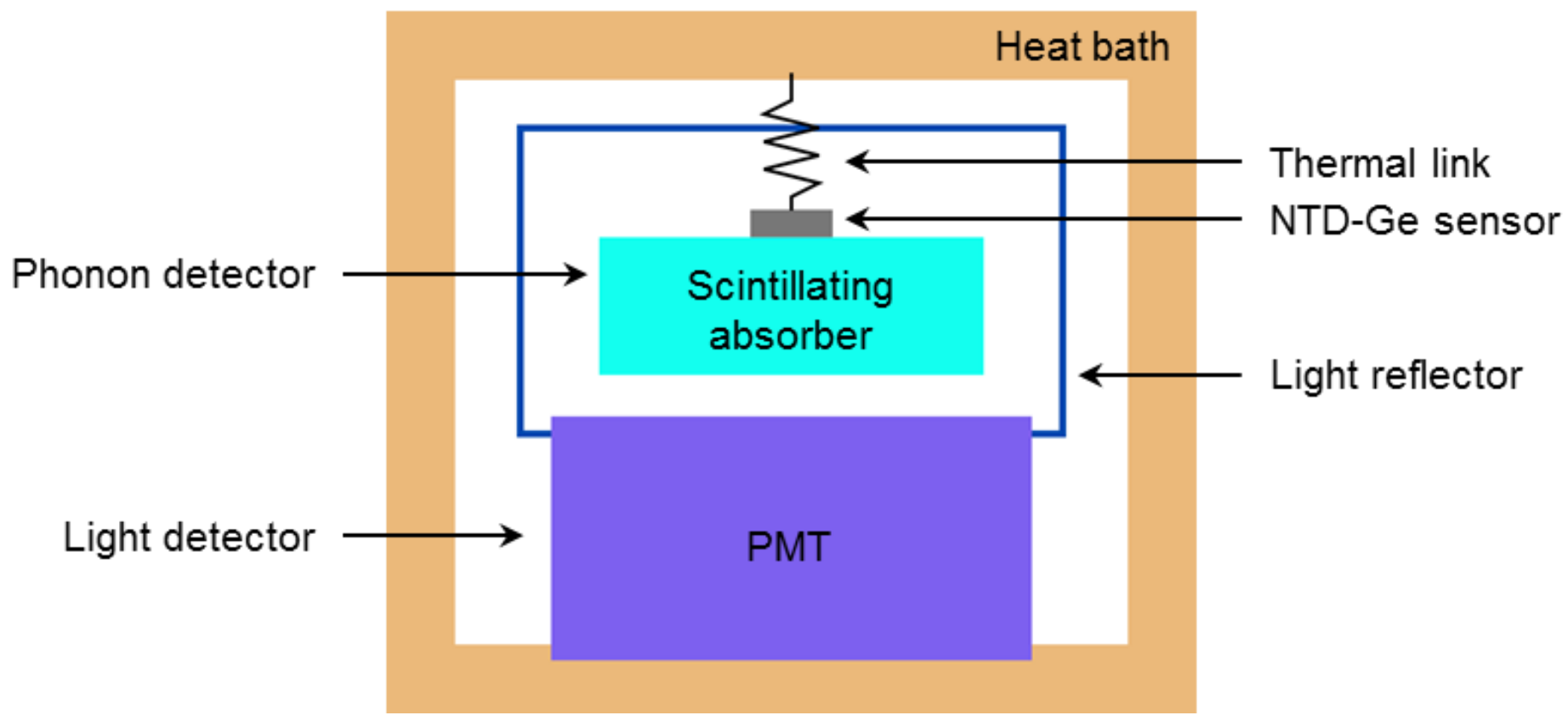

Fig. 1. Schematic view of the detector module.

### *3.1. Phonon detector*

The layout of the phonon detector is shown in Fig. 2. $CaWO_4$ and $CaMoO_4$ are used as the target crystal in the two detector modules tested in this study. A gold film is evaporated onto the top of the crystal to help collect phonons and thermalise them. The NTD-Ge sensor is glued onto this gold film with a small amount of Araldite adhesive. A Teflon fibre is placed in the adhesive between the NTD-Ge sensor and the gold film to prevent the NTD-Ge sensor from making direct contact with the gold film, which would electrically short the NTD-Ge sensor.

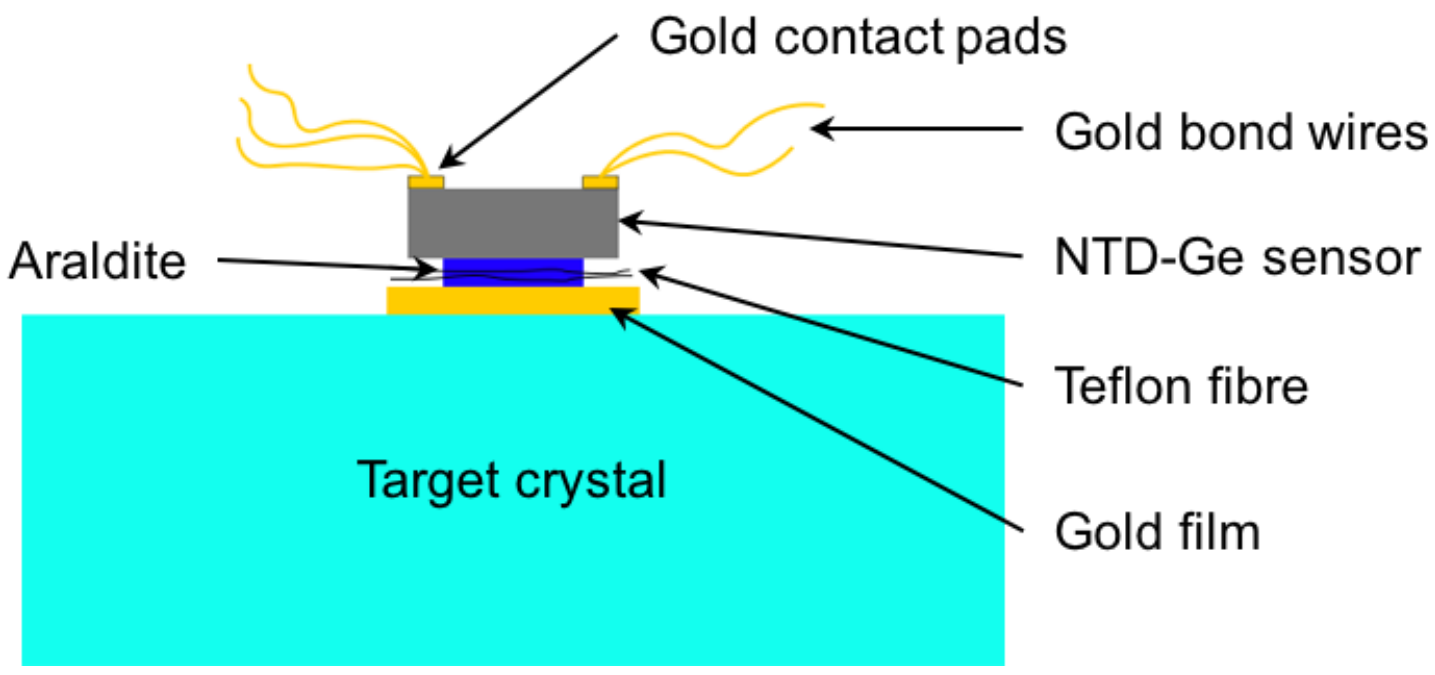

Fig. 2. Schematic of the phonon detector layout.

After a particle interaction, depositing energy in the target crystal, this energy eventually causes a temperature rise of the crystal, the gold film and the NTD-Ge sensor glued to it. This temperature rise is measured as a change of resistance of the sensor:

$$R(T) = R_0 \exp\left(\frac{T_0}{T}\right)^{1/2},$$

where $R_0$ and $T_0$ are constant and characteristic of the NTD-Ge sensor. Electric and thermal contact is provided through gold bond wires, linking gold contact pads on two opposite edges of the NTD-Ge sensor to landing pads on a small printed circuit board (PCB), mounted on the copper housing. The PCB has a receptacle array to provide connectivity for laminated, semi-flexible cabling, linking the NTD-Ge sensor to the front-end readout electronics.

*3.2. Light detector*

In this study, a Hamamatsu R8520-06 10-stage bialkali low-temperature PMT is used as the light detector. It faces the scintillating crystal without any contact. Since the PMT is positioned in the vicinity of the phonon detector, which is operated at milli-Kelvin temperatures and sensitive enough to pick up micro-Kelvin temperature changes, the power supply system that provides the high voltage (kV) to the PMT should not prevent the phonon detector from reaching its operating temperature or introduce noise to the heat channel signal. In this study, a high voltage system was built, based on a 25-stage Cockcroft–Walton generator (CWG) [43], installed at the Still level (~700 mK) of the dilution refrigerator used. The PMT is plugged into metal sockets mounted on a base PCB. Its cathode, dynodes and anode are then individually connected to different stages of the CWG through the PCB and a stainless steel laminated semi-flexible cable [44]. The PMT signal is carried by a coaxial cable all the way up to the readout electronics outside the cryostat.

To drive the CWG, a function generator positioned outside the cryostat provides an AC driving voltage (few V). This voltage is then amplified by a transformer (winding ratio 25:600) at 4.2 K and fed to the CWG through twisted wire pairs inside electrically grounded shielding. The high voltage generated is monitored externally after being stepped down by an attenuator (a resistive potential divider with resistors of room temperature values 1 MΩ and a 1 GΩ) mounted on the CWG PCB between the highest voltage stage and the monitoring cable. Fig. 3 shows the schematics of the circuit.

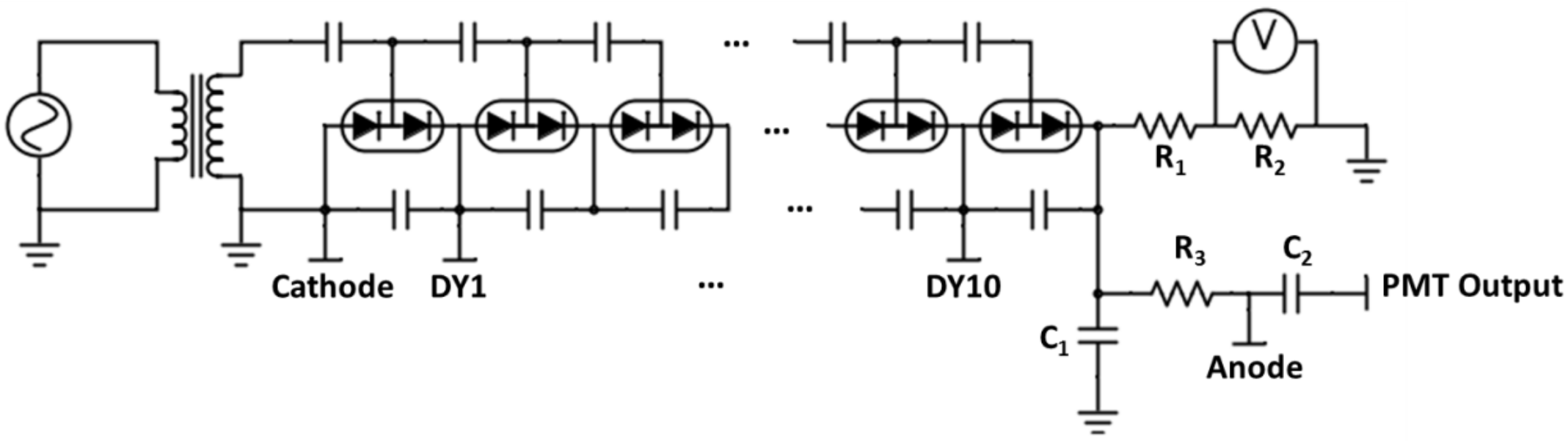

Fig. 3. Schematic of the high-voltage power supply, based on a 25-stage CWG design, and its connections to the PMT. Only the first and the last few stages of the CWG are shown for clarity. An attenuator consisting of two resistors, $R_1$ and $R_2$, for high-voltage monitoring is also shown (top right corner).

By using the capacitive CWG to deliver the various voltages directly to the cathode, anode and dynodes, instead of a resistive voltage divider, the heat dissipation from high-voltage generation is thus minimised, allowing the cryostat to cool to the required milli-Kelvin temperatures. Further, the high voltage is present only in the region of the CWG and the PMT, without need to supply high voltage from outside, thus eliminating high-voltage vacuum feedthroughs and long cryogenic high-voltage cabling.

No detrimental effect of running the CWG on operating a co-located, very sensitive low-temperature calorimeter has been found. The CWG is installed on the Still of the K400 cryostat, and for cryostat operation, the Still is equipped with a heater, the use of which is necessary to achieve the nominal cooling power the dilution refrigerator is capable of. The diodes and the dielectric losses in the capacitors of the CWG and the presence of the resistive attenuator generate a small amount of heat, which is less than usually supplied by the Still heater, and thus even beneficial. The usage of laminated semi-flexible cryogenic cabling [44] has been shown to be sufficient to shield the cryogenic calorimeter signal path from any adverse effect of the CWG operation, or passing 13 dynode / cathode / anode voltage supply wires in close proximity of the calorimeter readout might have. The CWG is operated at a frequency of 23 kHz, which is above the modulation frequency (1 kHz or 500 Hz) of the NTD-Ge sensor readout that we employed in the studies reported here, and it would also be above the typical bandwidth of SQUID-based readout systems used for transition-edge sensors [45]. The appearance of a single additional peak at 23 kHz can be identified in the frequency spectrum of the phonon detector signals, but its contribution to the overall noise power is so small that it does not result in a measurable increase of noise in the time domain.

### *3.3. Detector assembly*

Operating cryogenic detectors in a surface laboratory poses additional challenges compared to operation in the well-shielded environment inside an underground laboratory. The main issues are: natural radioactivity of the environment, cosmic rays and the muon flux that even in a small detector ($cm^3$ volume) causes pile-up due to the tens of milliseconds time constants of cryogenic detectors; vibration from an upper floor laboratory; and absence of a large thermal mass for temperature stabilisation, normally found in multi-detector module underground detectors. All of the above can be addressed by surrounding the actual detector with a large block of copper, cooled to the operating temperature of the actual detector. Fig. 4 shows the experiment setup, with three copper discs of 86 mm diameter, comprising the housing and thermal reservoir for the scintillating calorimeter and the low-temperature PMT.

A fourth disc of copper (74 mm diameter and 15 mm thickness) is installed below the PMT base to ensure sufficient shielding for the calorimeter on all sides. The compartment for the scintillating calorimeter with its NTD-Ge temperature sensor is lined with reflecting foil on five sides to maximize light collection and viewed by the PMT on its sixth face. The copper discs have one side of the round surface machined away to accommodate the PCB and connectors for the readout cabling, while opposite that, an opening is provided for radiation from the calibration source to pass through. After assembling, the prototype detectors (see Fig. 4) are mounted and tested in an Oxford Instruments K400 dilution refrigerator.

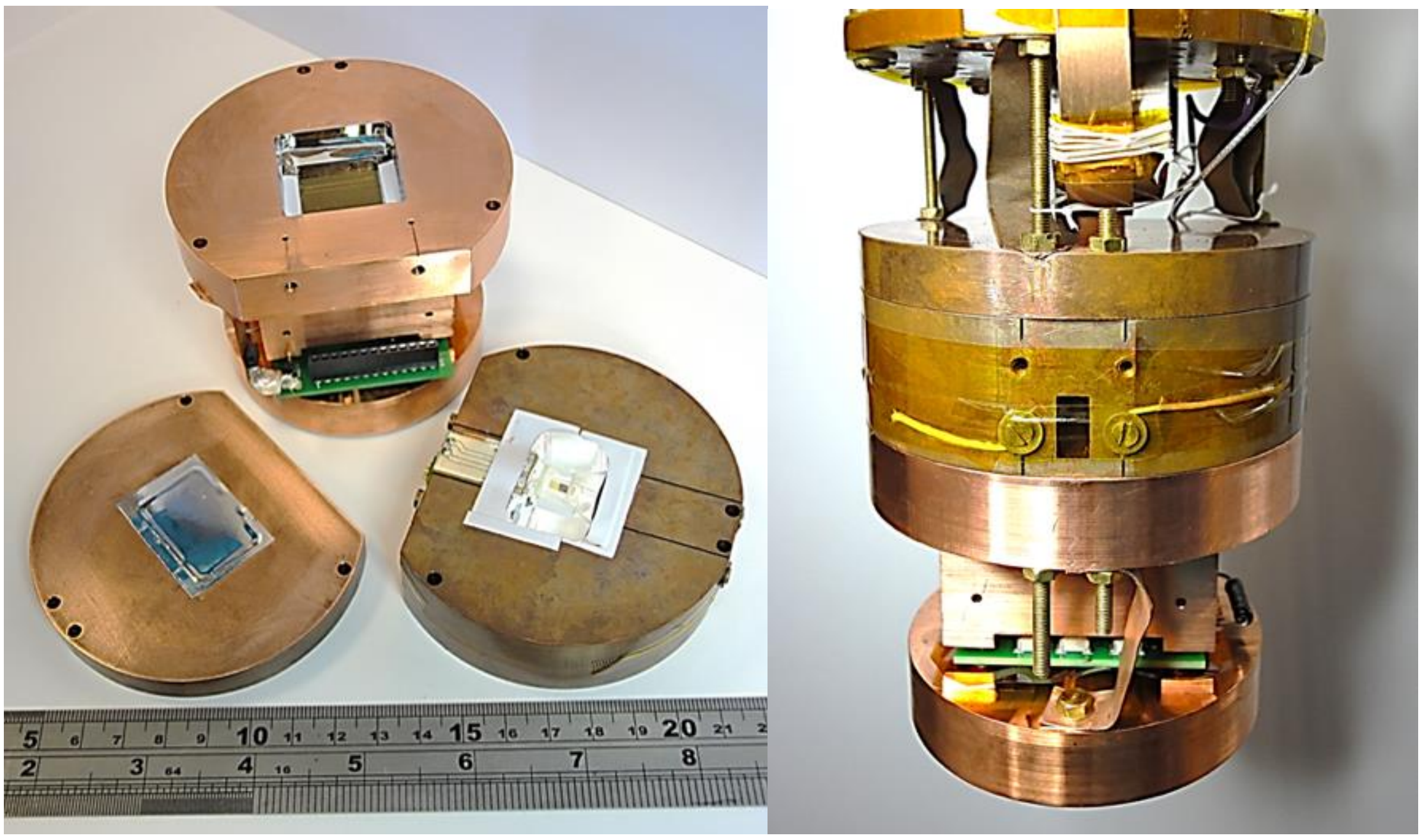

Fig. 4. Detector module assembly ($CaMoO_4$): (left) detector module components before final assembly: bottom left – top section with reflector closing off the $CaMoO_4$ calorimeter space; bottom right – middle section with the $CaMoO_4$ calorimeter suspended on strings within an opening through the copper block, lined with reflecting foil; and top – two copper pieces bolted together with the PMT base sandwiched between them and the PMT shrouded by one of the copper pieces; (right) complete detector module installed in the Oxford Instruments K400 dilution refrigerator – the PMT is facing up and the PMT base is visible just above the lowest of the copper pieces. The small opening between the tensioning screws for the calorimeter suspension allows irradiation of the calorimeter with gamma quanta from a calibration source.

### *3.4. Data acquisition*

The data from the light channel are collected using a fast custom-made data acquisition system. The PMT signal is digitized by an ADC with 5 ns sampling interval, thus allowing individual photons to be resolved. A 5 ns sampling interval means a clock frequency of 200 MHz, and implementing firmware on a XILINX Spartan-6 FPGA at that clock frequency throughout is challenging. A much better strategy is to confine this fast clock to a small fraction of the FPGA IO section and part of the fabric, through a scheme in which always 8 consecutive samples are grouped, thus relaxing the frequency requirement for the rest of the FPGA to a very comfortable 25 MHz. The groups of 8 samples continuously update the registers for storing the pre-trigger region (a ring buffer large enough to store the number of samples specified as pre-trigger region), from which the pre-trigger data are retrieved once the trigger condition is met. In parallel to storing the groups of eight data samples in the pre-trigger buffer, they are passed to the trigger logic in which the average of the eight samples in a block is calculated and used as value for evaluating whether the trigger condition is met. In the trigger mode applied here, the trigger condition is met when the 8-sample average changes to a value above a set threshold above baseline. This marks the start of the post-trigger region. Due to the 8-sample grouping, there is a maximum difference of 40 ns between the natural onset of the analogue signal pulse and the triggering point. Once triggered, the pre-trigger and post-trigger regions are propagated through an output FIFO and finally stored on a computer. This marks the end of a complete cycle of data acquisition, usually containing a single or few photons.

Two different modes of data taking were employed in this study. For $CaWO_4$ data, a fixed length of the overall data recording (pre- plus post-trigger region) was chosen. This method (example scintillation pulse shown in Fig. 14) produces a continuous record, capturing a scintillation event but has the disadvantage of introducing dead time while a large amount of data has to be read from the digitizer memory. For $CaMoO_4$ data, the trigger condition was modified such that the trigger remained set only until the 8-sample average dropped below the threshold for at least two 8-sample groups, at which point the end of the post-trigger region has been reached. In this way, only a small fraction of the baseline data in a scintillation event is recorded, and the decreased data rate enables "dead-time free" data acquisition. Typical scintillation events emit more than a dozen of photons within a short time interval, but detail depends on the decay time constant. In both cases, the analysis software extracts information on individual photons in the data set and groups them into scintillation events for which the number of photons detected is derived as well as photon arrival times which is then used to determine the time constants and scintillation amplitudes.

### *3.5. Light collection efficiency*

The main performance characteristics of CPSD, i.e. threshold, energy resolution, discrimination power as well as timing resolution, all depend on the number of detected photons which in turn is largely the product of three parameters: absolute scintillation light yield, quantum efficiency of photodetector and light collection efficiency of detection module. While the first two parameters are pretty much fixed for the specific selection of scintillation crystal and photodetector, it is possible to carry out optimization of the detection module’s light collection efficiency.

Light collection efficiency has recently been investigated for the specific configuration of CPSD, where no optical contact is permitted [46], [47]. It has been demonstrated that the light collection efficiency can change up to one-third depending on the geometry of detector, reflector and surface conditions of the scintillation crystal. These studies paved the way for optimisation of the light detection at cryogenic temperatures.

In the first test we used a polished rectangular $CaMoO_4$ scintillator crystal and obtained an acceptable energy resolution for scintillation detection. To improve the light channel response we built a second prototype detector module using $CaWO_4$ with roughened surfaces as the target crystal. The advantages of this configuration are twofold. On one hand, according to the manufacturer, the R8520-06 10-stage Bialkali PMT has higher quantum efficiency within the emission spectrum of $CaWO_4$ (broad band emission with a peak at 420 nm [18]) than $CaMoO_4$ (peak at 540 nm [18]). On the other hand, by roughening the surfaces of $CaWO_4$, the light collection efficiency of the detection module was maximised. We also modelled light collection for three geometries: A) control geometry, scintillator - PMT distance is 26 mm, B) scintillator - PMT distance reduced from 26 to 9 mm and C) the same as B but gap between back reflector and scintillator reduced to 3 mm. The input parameters for the simulations were taken from ref. [46]. The results of modelling indicated that reducing the distance between scintillator and PMT as well as moving the back reflector close to the crystal increases the efficiency of photon detection (see Fig. 5). This prediction was further confirmed by measuring the scintillation response of $CaWO_4$ at excitation with a $^{57}Co$ source at room temperature. Fig. 5 demonstrates how geometry optimisation leads to enhanced light collection efficiency of the scintillation module.

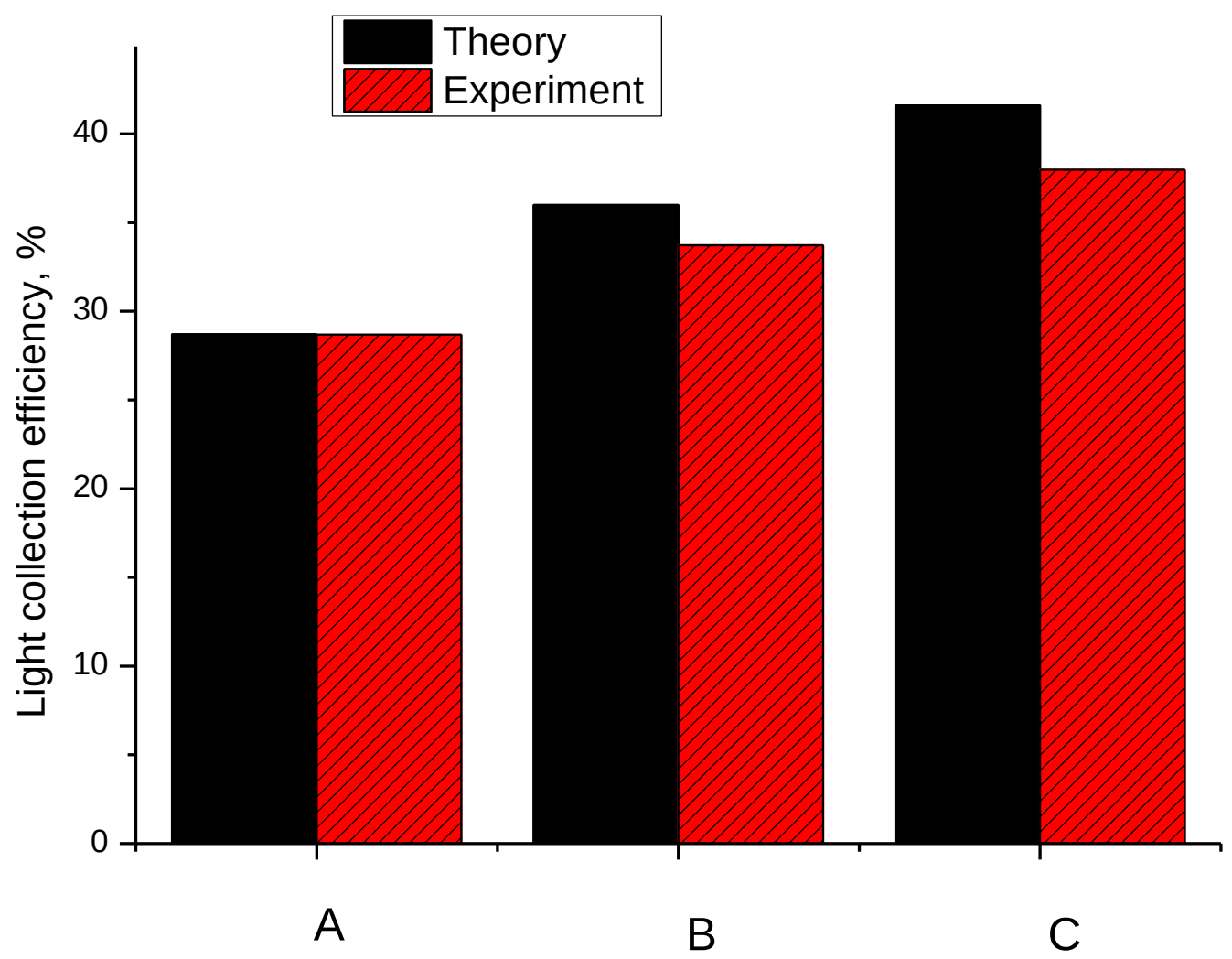

Fig. 5. Black bars - simulated light collection efficiency of detection module with $CaWO_4$ scintillator for different geometries A) control geometry, scintillator - PMT distance is 26 mm, B) scintillator - PMT distance reduced to 9 mm, C) the same as B but the gap between the back reflector and scintillator is reduced to 3 mm (from 19 mm in setup B). Red shaded bars display the scintillation response of the module measured at excitation with $^{57}Co$ source. For comparison, experimental results are normalised to the simulated value in control geometry A.

## 4. Results and discussion

### *4.1. Phonon channels*

The performance of the CPSDs was characterised using a $^{57}Co$ $\gamma$-source. Figures 6-9 show energy spectra obtained via the phonon and scintillation detector channels with $CaWO_4$ and $CaMoO_4$ scintillators as absorbers. The main features, the peaks at 122.1 and 136.5 keV, represent detection of the $\gamma$-quanta emitted by the $^{57}Co$ radioactive source (see Fig. 6 and 7) with the full energy remaining within the detector. Detecting less than the full energy can have a variety of reasons. Foremost, Compton scattering of the gammas off material surrounding the actual detector would result in a continuum of energies up to the full gamma energy. Or, Compton scattering can occur inside the detector, leading again to only partial detection of the full energy. There are also distinct line features that arise from the ionization of absorber nuclei, with, for example, a hole in the atom's K-shell. This hole will subsequently be filled by outer-shell electrons and the emission of a photon characteristic for the electron shell of the atom. If these photons leave the absorber, that energy is not detected. In the energy spectra for $CaWO_4$ this is visible as the escape peaks at 62.7 and 54.7 keV. These correspond to the loss of $K_\alpha$ and $K_\beta$ photons of tungsten, reducing the detected energy from the original 122.1 keV in $CaWO_4$ (Fig. 6).

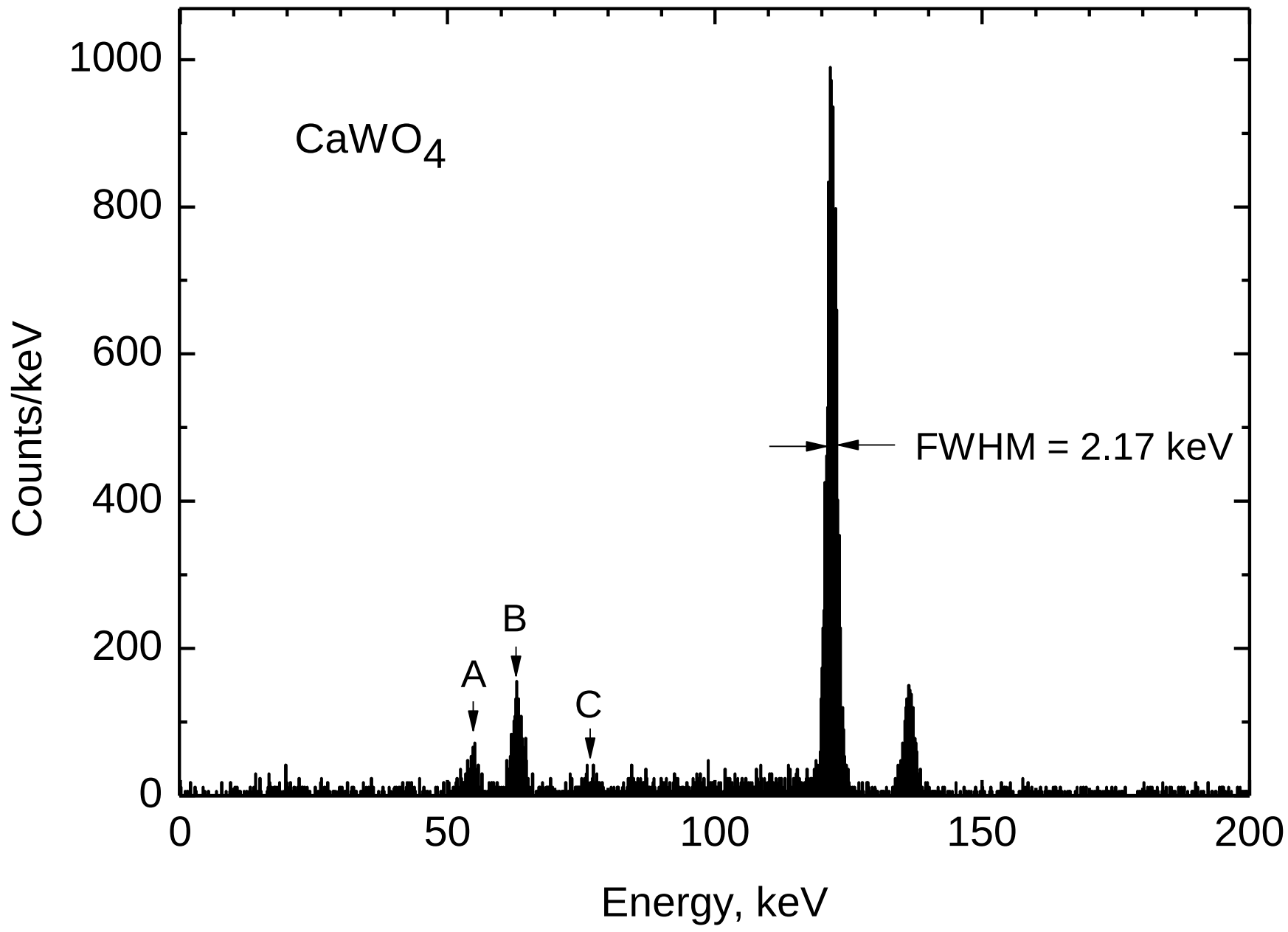

Fig. 6: Energy spectrum recorded by the phonon channel of the CPSD with a $CaWO_4$ absorber irradiated by a $^{57}$Co γ-source. The escape peaks A (122.1 keV – W-K$_\beta$), B (122.1 keV – W-K$_\alpha$), and C (136.5 keV – W-K$_\alpha$) are due to tungsten K$_\alpha$ and K$_\beta$ gamma having a significant probability of leaving the absorber crystal.

The same process will also occur for 136.5 keV gamma, but is less pronounced due to the smaller fraction of 136.5 keV gamma. An indication for K$_\alpha$ loss from these at around 77.2 keV is visible. Further features are more difficult to identify due to the continuum of other background. Molybdenum has a lower K-shell energy (molybdenum K$_\alpha$ and K$_\beta$ energies: 17.5 and 19.6 keV). These lower-energy photons have less probability to escape from the slightly larger absorber and thus the escape peaks in $CaMoO_4$ are less pronounced. Only a weak molybdenum K$_\alpha$ escape peak at 104.6 keV (Fig. 7) can be identified in $CaMoO_4$.

The energy resolution (FWHM) of the phonon channel at 122.1 keV is determined as 2.17 keV (1.8%) in $CaWO_4$ and 0.97 keV (0.79%) in $CaMoO_4$. The overall cut efficiency of the template fit is 49.7% for $CaWO_4$ and 52.7% for $CaMoO_4$. The energy resolution of the phonon channel observed in this study is inferior compared with 0.6 keV reported for a phonon detector that uses $CaWO_4$ absorber in a cryogenic Dark Matter search experiment [48]. On the other hand the energy resolution of measured CPSD with $CaMoO_4$ absorber is better than what has been achieved with magnetic metal calorimeter phonon sensors developed for the neutrinoless DBD experiment [49]. At this proof-of-principle stage we are far from having the optimal combination of all CPSD parameters as this was not a priority research aim. However, optimizing should not be too challenging, given the large number of examples of cryogenic detectors with excellent energy resolution in the phonon channel (see [3] and references therein).

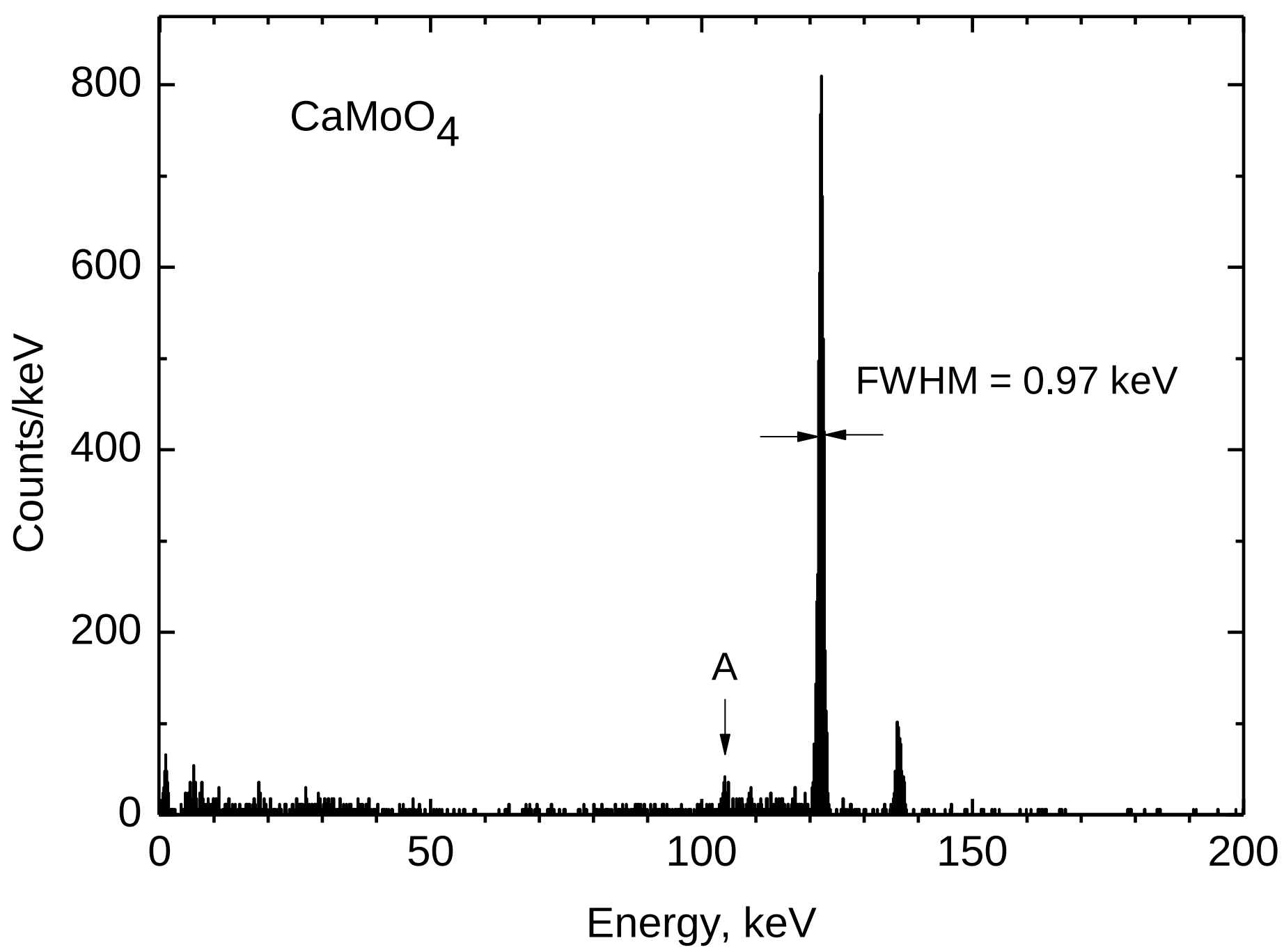

Fig. 7: Energy spectrum recorded by the phonon channel of the CPSD with a $CaMoO_4$ absorber irradiated by a $^{57}$Co γ-source. The escape peak A (122.1 keV – Mo-$K_\alpha$) is visible but less prominent than in tungsten due to molybdenum K-energies significantly lower and thus gamma less likely to leave the absorber crystal. Other escape peaks are too weak to be visible.

*4.2. Light channels*

The energy resolution of the light channel of CPSD is typically an order of magnitude worse than that of the phonon channel as only a few percent of the initial excitation energy contribute to a useful signal. With a PMT as the light detector in CPSD it is of particular interest to evaluate the quality of the measured scintillation response. The histogram of number of photons (N) detected in $CaWO_4$ and displayed in Fig. 8 exhibits two bands: the high-energy band with a peak at N=170 due to the absorption of 122.1 and 136.5 keV γ-rays of $^{57}$Co while the low-energy band is due to the escape lines of tungsten (the closely spaced individual lines cannot be resolved). The number of photons (N) was obtained by counting the number of single photon signals (see Fig. 14, top panel, for photon data). The energy resolution of the light detector at 122.1 keV is found to be 19.9%. By comparison, the best energy resolution of $CaWO_4$ at 122.1 keV (14.6%) has been measured at room temperature for a scintillator in optical contact with the PMT [50]. When the scintillator was used in CPSD applications, a value for the energy resolution equal to 17% has been achieved with a cryogenic light detector [22]. Nonetheless more typical values observed in Dark Matter search experiments are about 20% [51]. Thus, the light channel of our CPSD with $CaWO_4$ absorber exhibits a fairly competitive resolution in the low-energy range when compared with cryogenic calorimeter-based light detectors.

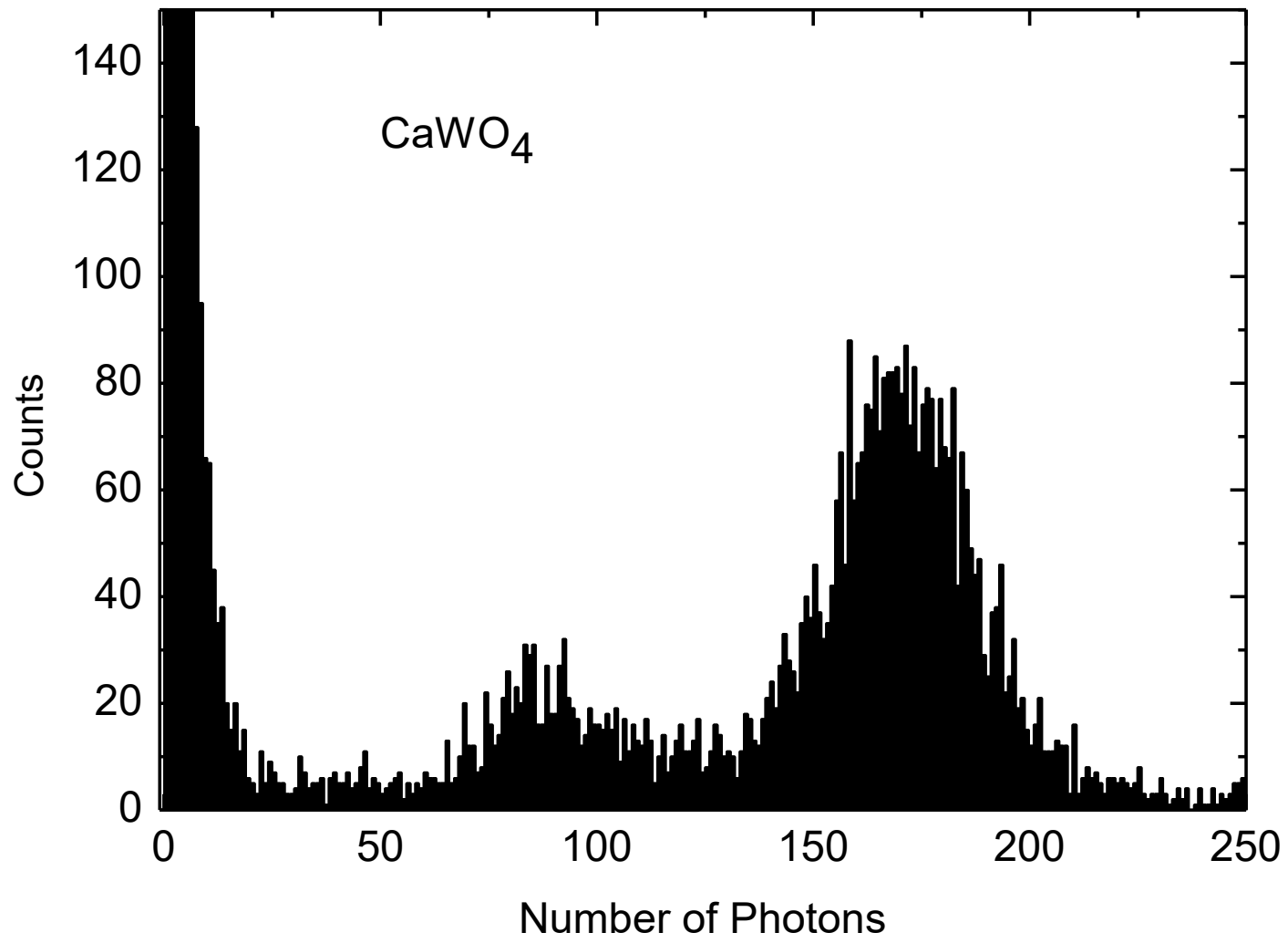

Fig. 8. Histogram of number of photons recorded by the light channel of the CPSD with a $CaWO_4$ scintillator irradiated by a $^{57}Co$ γ-source. The contributions to the histogram are the same as for the energy spectrum of the phonon channel.

Fig. 9 displays the energy spectrum detected with the $CaMoO_4$ crystal. The spectrum exhibits a broad band that is due to the interaction of $^{57}Co$ γ-rays with the scintillator. The observed energy resolution of the light channel is 29.7%. Although the absolute light yield of two crystals at cryogenic temperatures is fairly similar [18] the mean number of photons N=60 is significantly lower in $CaMoO_4$ due to the reduced spectral overlap of the scintillator emission spectrum and the PMT sensitivity curve [52] as well as a less efficient light collection of the first setup.

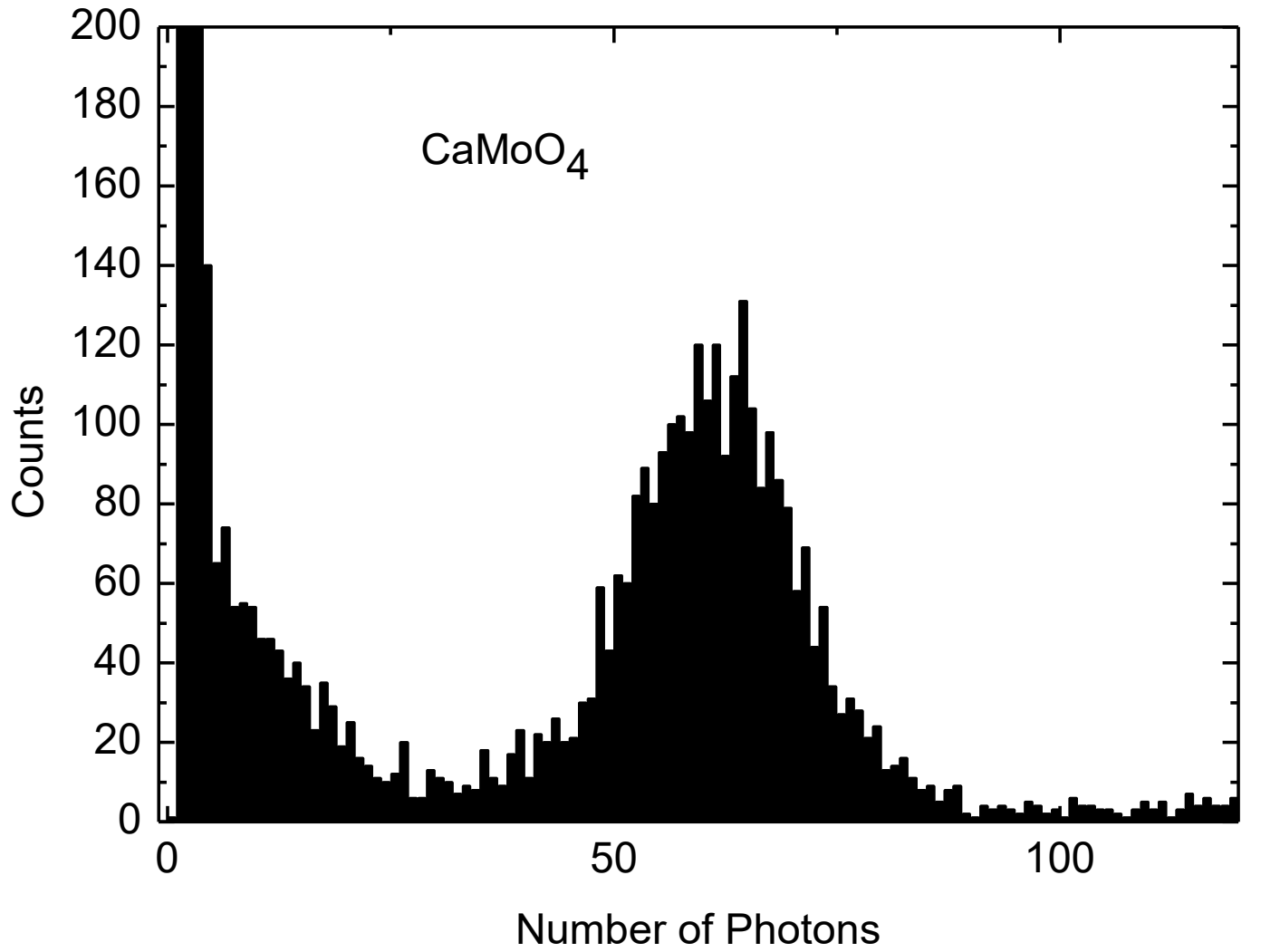

Fig. 9. Histogram of number of photons recorded by the light channel of the CPSD with a $CaMoO_4$ scintillator irradiated by a $^{57}$Co γ-source.

*4.3. Scintillation decay*

Fig. 10 shows the scintillation decay curves of $CaWO_4$ and $CaMoO_4$ scintillators as measured by the light channel of CPSD. The shape of the curves can essentially be characterized by two decay modes, fast and slow, which represent the relaxation of excited states in the scintillation material. The decay curves were fitted by the sum of two exponential functions with the fast and slow decay time constants listed in the Table 1.

Table 1. Parameters of decay kinetics in $CaWO_4$ and $CaMoO_4$ scintillators at 18 mK obtained from the fitting of the decay curves by the sum of two exponential functions: $y = A_f \exp\left(-\tau_f^{-1} t\right) + A_s \exp\left(-\tau_s^{-1} t\right) + y_0$.

| Crystal | $A_f$ | $\tau_f$ , μs | $A_s$ | $\tau_s$ , μs | $y_0$ |
|---|---|---|---|---|---|
| $CaWO_4$ | 460 | 12.8(6) | 2600 | 382(2) | 58 |
| $CaMoO_4$ | 2500 | 40.6 (8) | 158 | 3410(30) | 0 |

There is a striking difference between the two decay curves. First, the decay of $CaWO_4$ is controlled mainly by a slow component over the entire range, while the fast component is fairly feeble. In contrast, the fast component is dominant at the initial part of the $CaMoO_4$ decay curve; only when it vanishes, the slow component emerges. Second, the slow decay constant of $CaMoO_4$ is about an order of magnitude longer than in $CaWO_4$. Both these features can be explained in the framework of the model of emission centres, developed for tungstate and molybdate crystals [53], [54]. The energy scheme of the emission centre constitutes two excited levels of which the lower one is metastable. The levels are separated by a gap. Importantly, this gap in $CaMoO_4$ (0.61 meV) is almost one order of magnitude less than in $CaWO_4$. At cryogenic temperatures, when the phonon energy is comparable with the energy of this gap the kinetics of the radiative decay is governed by the dynamics of transitions between the two levels. In $CaWO_4$ the metastable level instantly accumulates all thermalized excitations and they decay predominantly from this level. Because of the smaller energy gap in $CaMoO_4$ there is a significant probability for the electrons to be promoted to the upper level from where they can undergo rapid radiative decay.

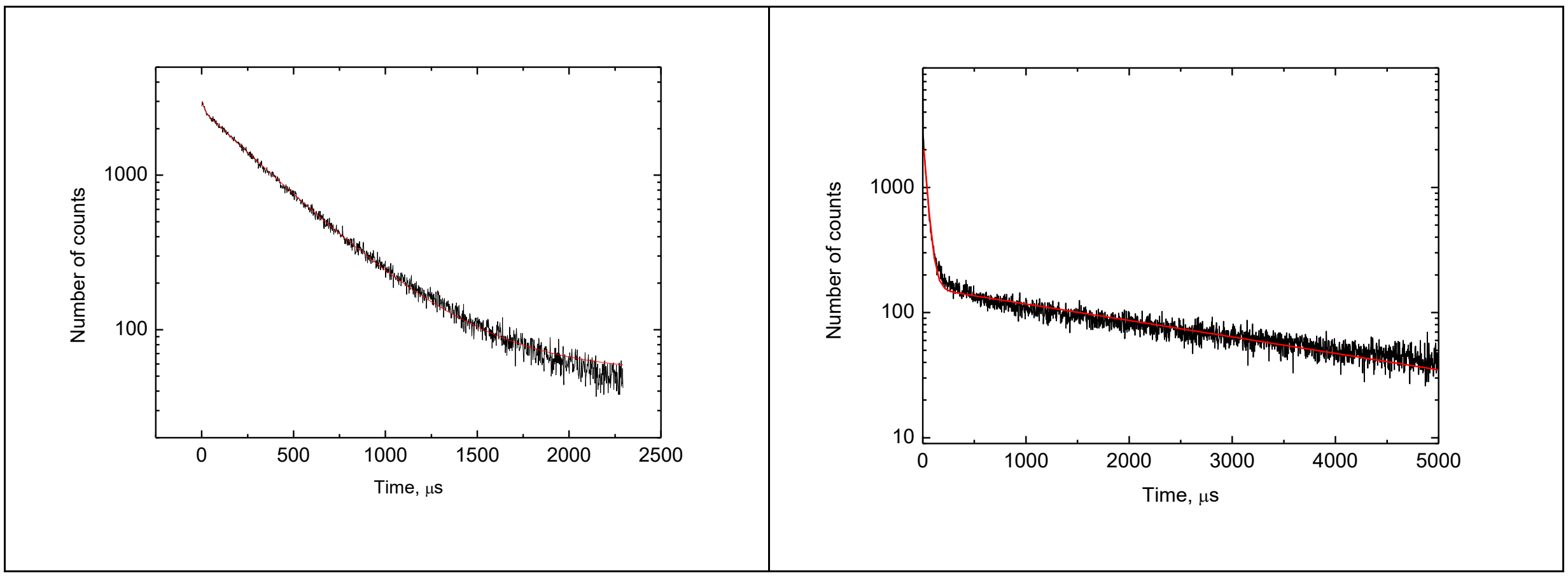

Fig. 10 Scintillation decay curves of $CaWO_4$ (left) and $CaMoO_4$ (right) measured at a temperature of 18 mK. The red lines show the best fit to the experimental data using the sum of two exponential functions: $y = A_f \exp\left(-\tau_f^{-1} t\right) + A_s \exp\left(-\tau_s^{-1} t\right) + y_0$.

*4.4. Simultaneous detection of phonon and light signals*

Next we assess the ability of CPSD to detect correlated signals from phonon and scintillation channel. The results displayed in Fig. 11 and 12 show a correlation plot of the signals detected by both channels. The horizontal axis shows the energy deposited in the phonon channel while the vertical is the number of photons measured by light detector. The diagonal band consists of events caused by gamma interactions which induce electron recoils. Massive particles and neutrons recoiling from nuclei yield much less light per unit of deposited energy [55] and subsequently these events will give rise to separate bands [5], [7], [20] (not visible here as the detectors were exposed to gamma quanta only). A different representation of the data shown in Fig. 11 ($CaWO_4$) is the yield plot (Fig. 13), in which the yield (ratio of photon signal versus phonon signal) is plotted against energy. The width of the observed band (for the gamma band, defined as yield 1), which is governed by the energy resolution of the light channel, provides a visual illustration for the discrimination ability of the CPSDs under test.

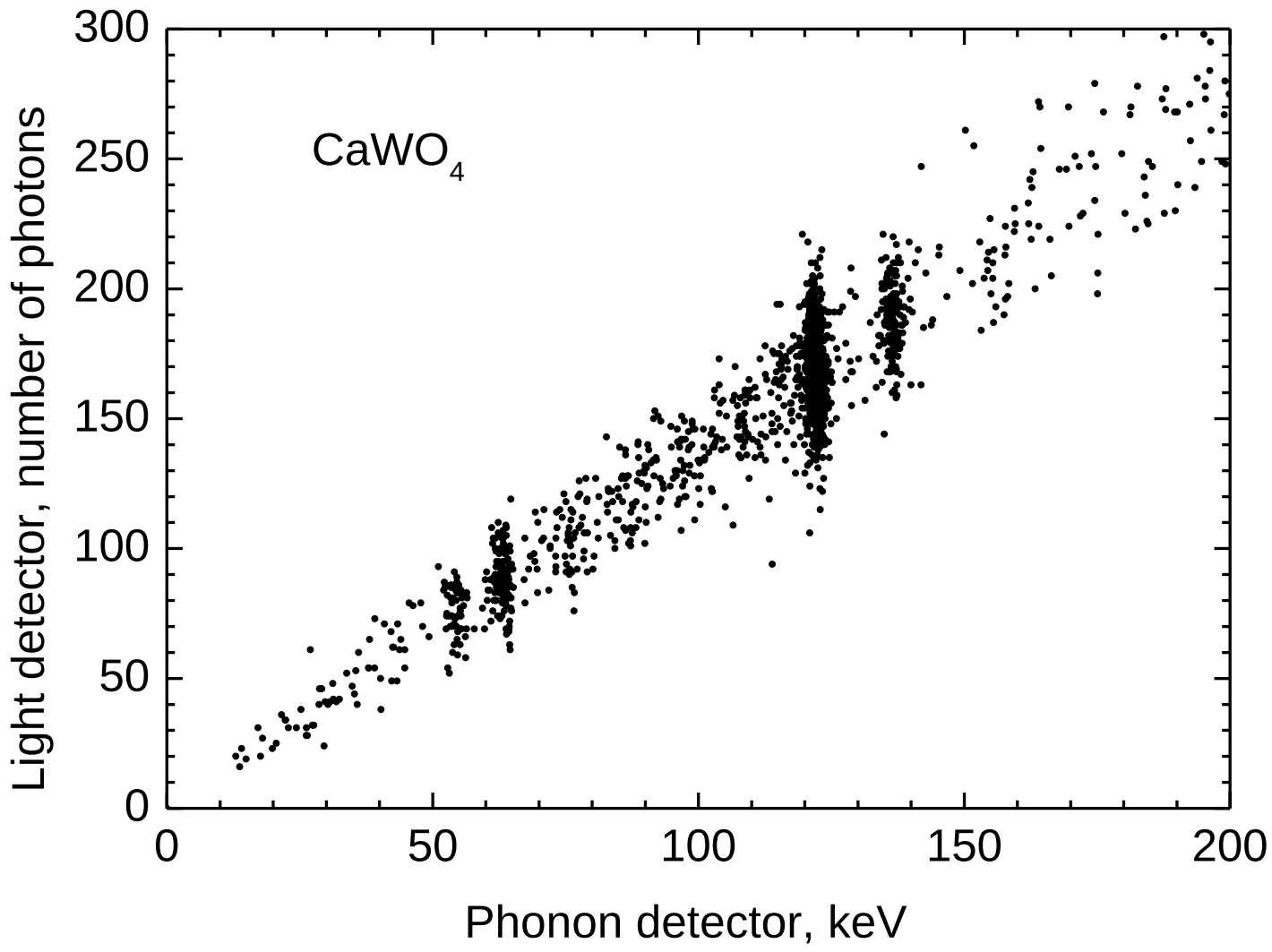

Fig. 11. Scatter plot of the energy in the light channel versus the energy in the phonon channel measured with a CPSD with $CaWO_4$ scintillator irradiated by a $^{57}$Co γ-source.

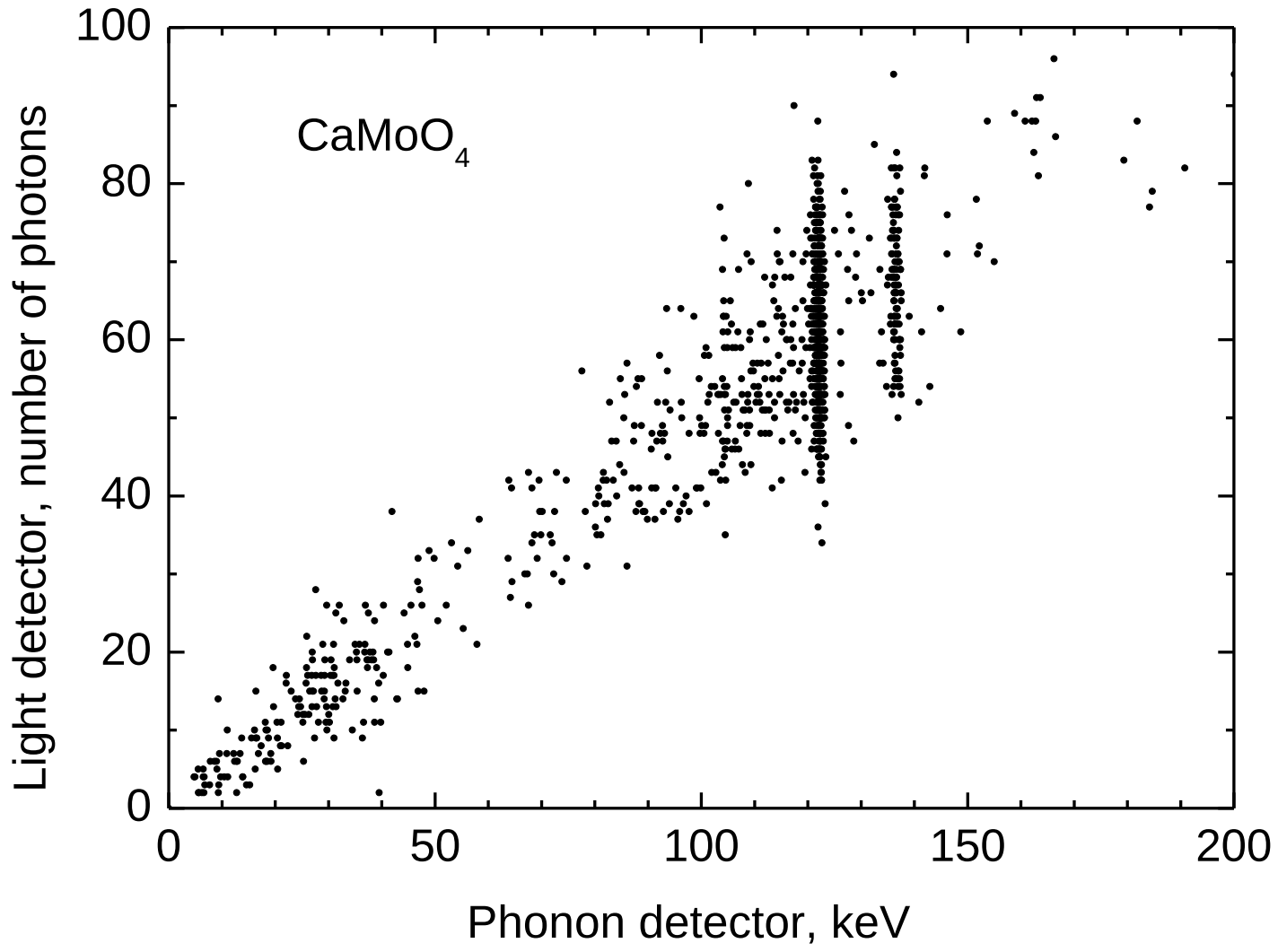

Fig. 12. Scatter plot of the energy in the light channel versus the energy in the phonon channel measured with a CPSD with $CaMoO_4$ scintillator irradiated by a $^{57}$Co γ-source.

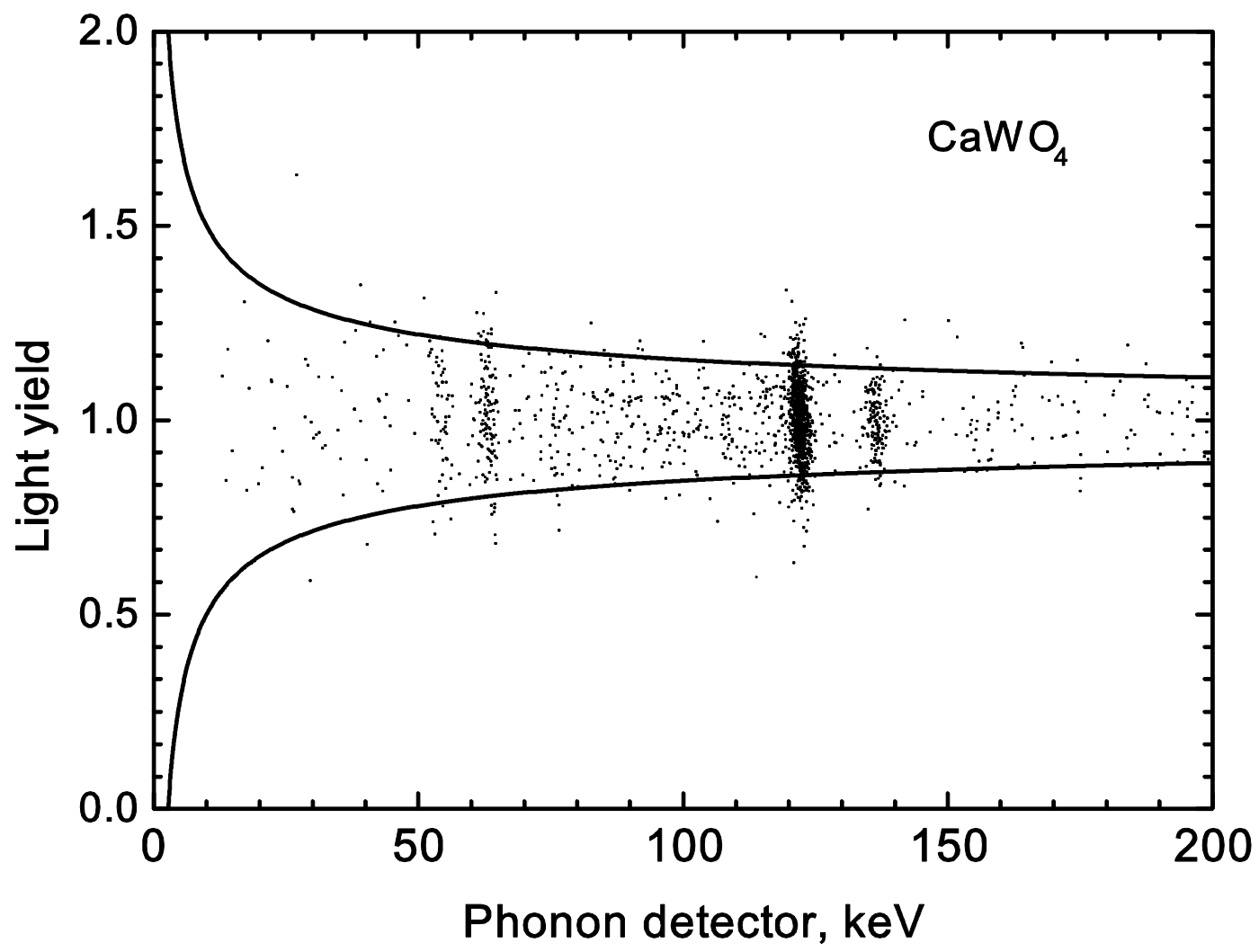

Fig. 13. Light yield versus energy of events. The solid lines indicate the 90% significance band of the electron recoil band at yield = 1.

*4.5. Pulse shape and time resolution*

Another important feature of the CPSD with PMT readout is the achievable timing resolution. This feature of the CPSD under study is illustrated in Fig. 14. The entire light signal is located at the very beginning of the phonon pulse. The time at which an interaction event occurred can be derived from the arrival of first photon in the scintillation pulse with a precision of a fraction of scintillation time constant (effectively the uncertainty between the interaction happening and the first detected photon emitted by the scintillator). This provides a well-defined trigger for time-correlated analysis of detected events and permits improving upon discrimination of cascade events and randomly coinciding events which are a cause of potential background for cryogenic experiments searching for neutrinoless DBD [56].

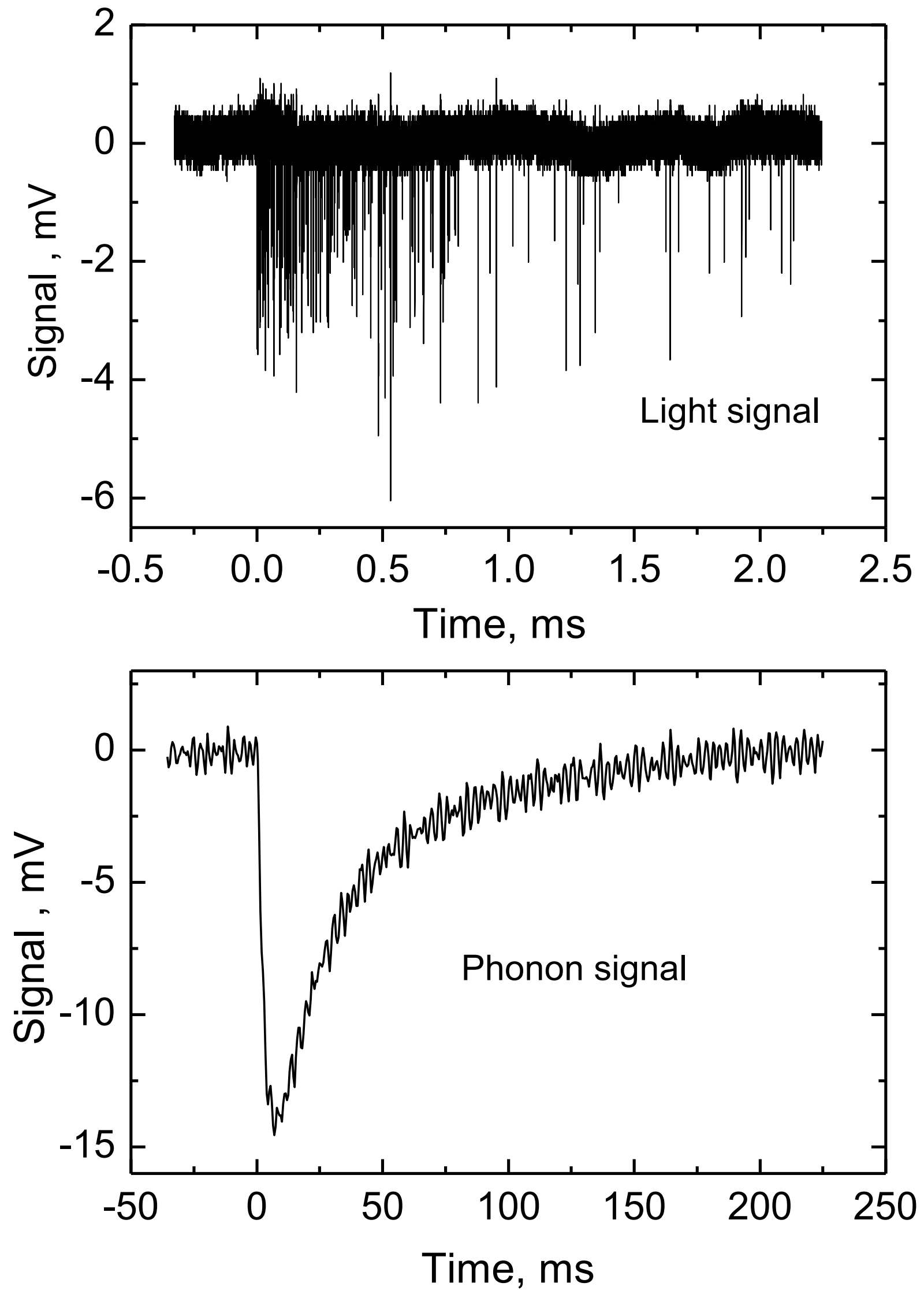

Fig. 14. Typical waveforms of the signals recorded by light (top) and phonon (bottom) channels of the CPSD following deposition of 122.1 keV γ-quanta in $CaWO_4$. Note a two order of magnitude difference of the time scale for the two plots.

Application of low-temperature PMT for detection of scintillations in CPSD offers new opportunities. Operating the detector module with the scintillation crystal at cryogenic temperatures permits harnessing a key feature of CPSD, i.e. excellent energy resolution and event-type discrimination. Moreover, the implementation of PMT as a light detector brings about a key advantage – ability of single photon detection that subsequently allows increasing the signal-to-noise ratio and also improves the timing characteristics of the detector. These two features are exceptionally useful for improving the CPSD capability for background identification and rejection. Furthermore scintillation signals detected by PMT can be utilised in a pulse shape analysis [57], providing thereby additional means of identification of different types of recoils at high energies.

*4.6 Intrinsic radioactivity*

Due to the extremely low rate of expected events, the sensitivity of the experiments is strongly affected by the radioactive background in and around the detector. Stringent requirements are imposed on the levels of intrinsic radioactivity of components used in the experimental setup [58]. Considerable effort is usually necessary to achieve the research aims regarding material selection [59], [60] purification [61] and production of detector components [62], [63] [64], aiming to reduce their intrinsic radioactivity levels. For the detector components themselves such requirements are obviously even more stringent. Therefore, it is essential to assess the contribution of the intrinsic radioactivity of PMT to the overall radioactive background of CPSD. Given currently achievable radiopurities, the natural benchmark in such assessment is given by the intrinsic radioactivity level of a scintillation crystal. Data on the intrinsic radioactivity of many materials are readily available in the literature; for example, the activity of radionuclides in a $CaWO_4$ crystal can be found in [64] and radiopurity results from several batches of Hamamatsu PMT model R8520-06 (same model but aluminium as underlay compared to platinum here) are presented in [59]. Recently, a calcium tungstate crystal of substantially improved radiopurity has been demonstrated [65]. These results have been obtained in situ with the $CaWO_4$ crystal operated as dark matter detector, while other radiopurity results were obtained through gamma spectroscopy.

In general, comparing the radiopurity results for $CaWO_4$ in [64] with PMT results in [59] suggests that scintillator and PMT are compatible, with the PMT radioactivity dominating only for $^{40}$K and $^{137}$Cs, both of which can be easily discriminated against due to the excellent electron/nuclear recoil rejection capability of cryogenic detectors. A real quantitative assessment on the individual impact of each contamination requires detailed background studies, taking into account all aspects of the experimental setup. Judging from the available data and taking into account that the PMT is an external component with respect to the volume of the scintillation crystal, it is reasonable to expect that the intrinsic radioactivity of the PMT will be only a minor contribution to the overall background. This is consistent with reports that limitation of sensitivity due to the residual counting rate of detectors in experiments equipped with CPSD is caused mainly by the intrinsic radioactivity of the scintillator material used as the target [48], [58], [66], [67]. PMT radioactivity does not appear to be a dominating factor, and if radiopurity levels of scintillator crystals improve, there are possibilities to reduce the effect of PMT radioactivity further, for example through using active scintillation shielding as suggested in [68].

## Conclusions

Cryogenic detectors capable of detecting simultaneously heat and light signals from particle interactions in a massive absorber may present a good choice for experiments searching for rare events. In this paper we investigated a CPSD module with a PMT as the detector for scintillation. The results of characterisations of two modules with $CaMoO_4$ and $CaWO_4$ scintillator respectively demonstrated that CPSD equipped with an NTD-Ge phonon sensor and a PMT as the detector of photons exhibits very promising performance characteristics. Data obtained with a $^{57}$Co source show that the detection module compares favourably with other CPSDs currently used in cryogenic rare-event search experiments. It retains the major advantages of conventional CPSD, i.e. high sensitivity, resolving power, and discrimination ability. The energy resolutions (FWHM) for excitation with 122.1 keV for the scintillation / PMT channel is 19.9% and 29.7% respectively for $CaWO_4$ and $CaMoO_4$ with the energy resolutions on the phonon channels 2.17 keV (1.8%) and 0.97 keV (0.79%).

The use of PMT offers significant improvement of time resolution of the CPSD. The fast response of PMT enables single-photon counting and as a result the time resolution of the detection module is governed merely by the slow scintillation decay time constant of the crystal used. These were found to be 382±2 μs in $CaWO_4$ and 3410±30 μs in $CaMoO_4$. Furthermore, as the shape of the scintillation response recorded by the PMT depends on the ionization density, pulse shape analysis can be used for separation between electron and nuclear recoils, offering additional possibilities for background identification and rejection.

In summary, we presented here the novel concept of a cryogenic phonon-scintillation detector, equipped with PMT readout and PMT dynode voltage generation by a Cockcroft-Walton generator at low temperature, and demonstrated that it is a very promising cryogenic detector for future cryogenic rare-event search experiments.

**Acknowledgment**

The study was supported in part by the Science & Technology Facilities Council (STFC). The authors thank the EDELWEISS collaboration, in particular Xavier-Francois Navick (CEA, Saclay), for providing the NTD-Ge sensor used in this study. XZ and JL thank the China Scholarship Council for their support.

## References

[1] A.H. G. Peter, V. Gluscevic, A. M. Green, B. J. Kavanagh and S. K. Lee , "WIMP physics with ensembles of direct-detection experiments," *Physics of the Dark Universe,* Vols. 5-6, pp. 45-74, 2014.

[2] S. R. Elliott, "Recent Progress in Double Beta Decay," *Mod. Phys. Lett. A*, vol. 27, p. 1230009, 2012.

[3] C. Enss and D. McCammon , "Physical principles of low temperature detectors: ultimate performance limits and current detector capabilities," *J. Low. Temp. Phys.* , vol. 151, pp. 5-24, 2008.

[4] A. Alessandrello, V. Bashkirov, C. Brofferio, C. Bucci, D.V. Camin, O. Cremonesi, E. Fiorini, G Gervasio, A. Giuliani, A. Nucciotti, M. Pavan, G. Pessina, E. Previtali and L. Zanotti, "A scintillating bolometer for experiments on double beta decay," *Phys. Lett. B,* vol. 420, pp. 109-113, 1998.

[5] P. Meunier, M. Bravin, M. Bruckmayer, S. Giordano, M. Loidl, O. Meier, F. Pröbst, W. Seidel, M. Sisti, L. Stodolsky et al, "Discrimination between nuclear recoils and electron recoils by simultaneous detection of phonons and scintillation light," *Appl. Phys. Lett.,* vol. 75, pp. 1335-1337, 1999.

[6] V. B. Mikhailik and H. Kraus, "Cryogenic scintillators in searches for extremely rare events," *Journal of Physics D: Applied Physics,* vol. 39, pp. 1181-1191, 2006.

[7] R.F. Lang, G. Angloher, M. Bauer, I. Bavykina, A. Bento, A. Brown, C. Bucci, C. Ciemniak, C. Coppi, G. Deuter, et al., "Discrimination of recoil background in scintillating calorimeters," *Astroparticle Physics,* vol. 33, pp. 60-64, 2010.

[8] P. de Marcillac, N. Coron, G. Dambier, J. Leblanc and J. -P. Moalic , “Experimental detection of alpha-particles from the radioactive decay of natural bismuth,” *Nature,* vol. 422, pp. 876-878 , 2003.

[9] C. Cozzini, G. Angloher, C. Bucci, F. von Feilitzsch, D. Hauff, S. Henry, Th. Jagemann, J. Jochum, H. Kraus, B. Majorovits et al., “Detection of the natural a-decay of tungsten,” *Phys. Rev.C* , vol. 70, p. 064606, 2004.

[10] N. Casali, S. S. Nagorny, F. Orio, L. Pattavina, J. W. Beeman, F. Bellini, L. Cardani, I. Dafinei, S. Di Domizio, M. L. Di. Vacri et al.,, “Discovery of the 151Eu α decay,” *J. Phys. G: Nucl. Part. Phys.,* vol. 41, p. 075101, 2014.

[11] S. Roth, G. Angloher, M. Bauer, I. Bavykina, A. Bento, A. Brown, C. Bucci, C. Ciemniak, C. Coppi, G. Deuter, et al., “Direct dark matter search with CRESST and EURECA,” *Progress in Particle and Nuclear Physics,* vol. 64, pp. 457-459, 2010.

[12] A. Brown, S. Henry, H.Kraus and C. McCabe, “Extending the CRESST-II commissioning run limits to lower masses,” *Physical Review D,* vol. 85, pp. 021301(R)-5, 2012.

[13] G. Angloher, A. Bento, C. Bucci, L. Canonica, A. Erb, F. von Feilitzsch, N.F. Iachellini, P. Gorla, A. Guetlein, et al., “Results on low mass WIMPs using an upgraded CRESST-II detector,” *Eur. Phys. J. C* , vol. 3184 , p. 74, 2014.

[14] L. Gironi, “Scintillating bolometers for Double Beta Decay search,” *Nuclear Instruments and Methods in Physics Research A,* vol. 617, pp. 478-481, 2010.

[15] J. W. Beeman, F. A.Danevich, V. Ya. Degoda, E.N. Galashov, A. Giuliani, V.V. Kobychev, M. Mancuso, S. Marnieros, C. Nones, E. Olivieri et al., “A next-generation neutrinoless double beta decay experiment based on ZnMoO4 scintillating bolometers,” *Phys. Lett. B,* vol. 710, pp. 318-323, 2012.

[16] E. Armengaud, Q. Arnaud, C. Augier, A. Benoît, A. Benoît, L. Bergé, R.S. Boiko, T. Bergmann, J. Blümer, A. Broniatowski et al., “Development and underground test of radiopure ZnMoO4 scintillating bolometers for the LUMINEU 0ν2β project,” *Journal of Instrumentation,* vol. 10, p. P05007, 2015.

[17] G. B. Kim, S. Choi, F. A. Danevich, A. Fleischmann, C. S. Kang, H. J. Kim, S. R. Kim, Y. D. Kim, Y. H. Kim, V. A. Kornoukhov et al., “A CaMoO4 Crystal Low Temperature Detector for the AMoRE Neutrinoless Double Beta Decay Search,” *Adv. High Ener. Phys.* , vol. 2015 , p. 817530, 2015 .

[18] V. B. Mikhailik and H. Kraus, “Performance of scintillation materials at cryogenic temperatures,” *Phys. Stat. Sol. B,* vol. 247, pp. 1583-1599, 2010.

[19] C. Arnaboldi, J.W. Beeman, O. Cremonesi, L. Gironi, M. Pavan, G. Pessina, S. Pirro, E. Previtali , “CdWO4 scintillating bolometer for Double Beta Decay: Light and heat anticorrelation, light yield and quenching factors,” *Astroparticle Physics* , vol. 34, p. 143–150, 2010.

[20] N. Coron, P. de Marcillac, J. Leblanc, G. Dambier and J.-P. Moalic , “Highly sensitive large-area bolometers for scintillation studies below 100 mK,” *Opt. Eng.,* vol. 47, p. 1568–1576 , 2004.

[21] D. Rosenberg, A. E. Lita, A. J. Miller, and S. W. Nam, “Noise-free high-efficiency photon-number-resolving detectors,” *Physics Review A,* vol. 71, p. 061803R, 2005.

[22] F. Petricca, G. Angloher, C. Cozzini, T. Frank, D. Hauff, J. Ninkovic, F. Probst, W. Seidel, S. Uchaikin , “Light detector development for CRESST II,” *Nuclear Instruments and Methods in Physics Research A* , vol. 520, p. 193–196, 2004.

[23] G. Angloher, M. Bauer, I. Bavykina, A. Bento, A. Brown, C. Bucci, C. Ciemniak, C. Coppi, G. Deuter, F. von Feilitzsch et al., “Commissioning run of the CRESST-II dark matter search,” *Astropart. Phys.,* vol. 31, pp. 270-276, 2009.

[24] P. C. F. Di Stefano, T. Frank, G. Angloher, M. Bruckmayer, C. Cozzini, D. Hauff, F. Probst, S. Rutzinger, W. Seidel, and L. Stodolsky , “Textured silicon calorimetric light detector,” *J. Appl. Phys.* , vol. 94, pp. 6887-6891, 2003.

[25] A. Calleja, N. Coron, E. García, J. Gironnet, J. Leblanc, P. de Marcillac, M. Martínez, Y. Ortigoza, A. Ortiz de Solórzano, C. Pobes et al. , “Recent performance of scintillating bolometers developed for Dark Matter searches,” *J. Low Temp. Phys.* , vol. 151, p. 848–853, 2008.

[26] J.W. Beeman, F. Bellini, N. Casali, L. Cardani, I. Dafinei, S. Di Domizio, F. Ferroni, L. Gironi, S. Nagorny et al. , “Characterization of bolometric light detectors for rare event searches,” *J. Instrum* , vol. 8, p. P07021, 2013.

[27] C. Isaila, C. Ciemniak, F.v. Feilitzsch, A. Gütlein, J. Kemmer, T. Lachenmaier, J.-C. Lanfranchi, S. Pfister, W. Potzel, S. Roth et al., “Low-temperature light detectors: Neganov–Luke amplification and calibration,” *Phys. Let. B* , vol. 716, p. 160–164, 2012.

[28] L. Pattavina, N. Casali, L. Dumoulin, A. Giuliani, M. Mancuso, P. de Marcillac, S. Marnieros, S. S. Nagorny, C. Nones, E. Olivieri, L. Pagnanini, S. Pirro, D. Poda, C. Rusconi, K. Scheffner, M. Tenconi, “Background Suppression in Massive TeO2 Bolometers with Neganov–Luke Amplified Light Detectors,” *J.Low Temp. Phys.* , vol. 181, pp. 1-6, 2015.

[29] H.J. Lee, J.H.So, C.S.Kang, G.B.Kim, S.R.Kim, J.H.Lee , M.K.Lee , W.S.Yoon, Y.H.Kim, “Development of a scintillation light detector for a cryogenic rare-event-

search experiment," *Nuclear Instruments and Methods in Physics Research A,* vol. 784, p. 508–512, 2015.

[30] L. Cardani, I. Colantoni, A. Cruciani, S. Di Domizio, M. Vignati, F. Bellini, N. Casali,M. G. Castellano, A. Coppolecchia, C. Cosmelli and C. Tomei, "Energy resolution and efficiency of phonon-mediated kinetic inductance detectors for light detection," *Applied Physics Lerrers* , vol. 107, p. 093508 , 2015.

[31] Yang, S.N. Dzhosyuk, J.M. Gabrielse, P.R. Huffman, C.E.H. Mattoni, S.E. Maxwell, D.N. McKinsey and J.M. Doyle, "Performance of a large-area avalanche photodiode at low temperature for scintillation detection," *Nuclear Instruments and Methods in Physics Research A,* vol. 508, pp. 388-393, 2003.

[32] G. Collazuol, M. G. Bisogni, S. Marcatili, C. Piemonte and A. DelGuerra, "Studies of silicon photomultipliers at cryogenic temperatures," *Nuclear Instruments and Methods in Physics Research A,* vol. 628, p. 389–392, 2011.

[33] M. Biroth, P.Achenbach, E.Downie and A.Thomas, "Silicon photomultiplier properties at cryogenic temperatures," *Nuclear Instruments and Methods in Physics Research A,* vol. 787, p. 68–71, 2015.

[34] A. Bueno, J. Lozano, A. J. Malgarego, F. J. Munoz, J. L. Navarro, S. Navas and A. G. Ruiz, "Characterization of large area photomultipliers and its application to dark matter search with noble liquid gases," *Journal of Instrumentation,* vol. 3, p. P01006, 2008.

[35] F. Carbanora, A.G.Cocco, G.Fiorillo and V.S.Gallo, "Performance of photomultiplier tubes for cryogenic applications," *Nuclear Instruments and Methods in Physics Research A,* vol. 610, pp. 271-275, 2009.

[36] A. Falcone ,R. Bertonic, F. Boffelli, M. Bonesini, T. Cervi, A. Menegolli, C. Montanari, M.C. Prata, A. Rappoldi, G. L. Raselli et al., "Comparison between large area photo-multiplier tubes at cryogenic temperature for neutrino and rare event physics experiments," *Nuclear Instruments and Methods in Physics Research A,* vol. 787, pp. 55-58, 2015.

[37] L. Baudis, "WIMP dark matter direct-detection searches in noble gases," *Physics of the Dark Universe,* vol. 4, pp. 50-59, 2014.

[38] H. Kraus and V. B. Mikhailik , "First test of a cryogenic scintillation module with a CaWO4 scintillator and a low-temperature photomultiplier down to 6 K," *Nuclear Instruments Methods Physics Researches A* , vol. 621, pp. 395-400, 2010.

[39] H.O. Meyer, "Performance of a photomultiplier at liquid-helium temperature," *Nuclear Instruments and Methods in Physics Research A,* vol. 621, p. 437–442, 2010.

[40] T. M. Ito, S. M. Clayton, J. Ramsey, M. Karcz, C.-Y. Liu, J. C. Long, T. G. Reddy, and G. M. Seidel, "Effect of an electric field on superfluid helium scintillation produced by α-particle sources," *Physical Review A,* vol. 85, p. 042718, 2012.

[41] G S Buller and R J Collins, "Single-photon generation and detection," *Meas. Sci. Technol.,* vol. 21 , p. 012002 , 2010.

[42] M. F. Weber, C. A. Stover, L. R. Gilbert, T. J. Nevitt and A. J. Ouderkirk , "Giant Birefringent Optics in Multilayer Polymer Mirrors," *Science* , vol. 287, pp. 2451-2456 , 2000.

[43] J. D. Cockcroft and E. T. S. Walton , "Experiments with high velocity positive ions.(I) Further developments in the method of obtaining high velocity positive ions," *Proceedings of the Royal Society A,* vol. 136, p. 619–630, 1932.

[44] X. Zhang , "A novel phonon-scintillation cryogenic detector and cabling solution for Dark Matter direct detection, D.Phil thesis," University of Oxford, 2015.

[45] S.Henry, N.Bazin, H.Kraus, B.Majorovits, M.Malek, R.McGowan, V.B.Mikhailik, Y.Ramachers and A.J.B.Tolhurst, "The 66-channel SQUID readout for CRESST II," *journal of Instrumentation,* vol. 2, p. P11003, 2007.

[46] F.A. Danevich, R. V. Kobychev, V. V. Kobychev et al., "Impact of geometry on light collection efficiency of scintillation detectors for cryogenic rare event searches," *Nuclear Instruments Methods Physics Researches B,* vol. 336, pp. 26-30, 2014.

[47] F.A. Danevich, R. V. Kobychev, V. V. Kobychev, H. Kraus, V. B. Mikhailik, and V. M. Mokina, "Optimization of light collection from crystals scintillators for cryogenic experiments," *Nuclear Instruments and Methods A,* vol. 744, pp. 41-47, 2014.

[48] R.F. Lang, G. Angloher, M. Bauer, I. Bavykina, A. Bento, A. Brown, C. Bucci, C. Ciemniak, C. Coppi, G. Deuter, et al., "Electron and gamma background in CRESST detectors," *Astroparticle Physics,* pp. 318-324, 2010.

[49] S.J. Lee, J.H. Choi, F.A. Danevich, Y.S. Jang, W.G. Kang, N. Khanbekov, H.J. Kim, I.H. Kim, S.C. Kim, S.K. Kim et al. , "The development of a cryogenic detector with CaMoO4 crystals for neutrinoless double beta decay search," *Astroparticle Physics,* vol. 34, p. 732–737, 2011.

[50] M. Moszyński, M. Balcerzyk, W. Czarnacki, A. Nassalski, T. Szczęśniak, H. Kraus, V. B. Mikhailik and I. M. Solskii, "Characterization of CaWO4 scintillator at room and liquid nitrogen temperatures," *Nuclear Instruments and Methods in Physics Research A,* vol. 553, pp. 578-591, 2005.

[51] R. F. Lang and W. Seidel , "Search for dark matter with CRESST," *New Journal of*

*Physics,* vol. 11, p. 105017, 2009.

[52] V. B. Mikhailik, S. Henry, H. Kraus and I. Solskii, “Temperature dependence of CaMoO4 scintillation properties,” *Nuclear Instruments and Methods in Physics Research A,* vol. 583, pp. 350-355, 2007.

[53] V. B. Mikhailik, H. Kraus, S. Henry and A. J. B. Tolhurst, “Scintillation studies of CaWO4 in the millikelvin temperature range,” *Phys. Rev. B,* vol. 75, p. 184308, 2007.

[54] X. Zhang, J. Lin, V. B. Mikhailik and H. Kraus , “Studies of scintillation properties of CaMoO4 at millikelvin temperatures,” *Applied Physics Letters* , vol. 106, p. 241904, 2015.

[55] V.I. Tretyak, “Semi-empirical calculation of quenching factors for ions in scintillators,” *Astroparticle Physics,* vol. 33, pp. 40-53 , 2010.

[56] D. M. Chernyak, F. A. Danevich, A. Giuliani et al., “Rejection of randomly coinciding events in ZnMoO4 scintillating bolometer,” *Phys. J. C* , vol. 74, p. 2913, 2014.

[57] L. Bardelli, M. Bini, P.G. Bizzeti, L. Carraresi, F.A. Danevich, T.F. Fazzini, B.V. Grinyov, N.V. Ivannikova, V.V. Kobychev, B.N. Kropivyansky, P.R. Maurenzig, L.L. Nagornaya, S.S. Nagorny, A.S. Nikolaiko, A.A. Pavlyuk, D.V. Poda et al.,, “Further study of CdWO4 crystal scintillators as detectors for high sensitivity 2b experiments: Scintillation properties and pulse-shape discrimination,” *Nuclear Instruments and Methods in Physics Research A,* vol. 569, p. 743–753, 2006.

[58] V. Tomasello, M. Robinson, V.A. Kudryavtsev , “Radioactive background in a cryogenic dark matter experiment,” *Astroparticle Physics* , vol. 34, p. 70–79, 2010.

[59] E. Aprile, K. Arisaka, F. Arneodo, A. Askin, L. Baudis, A. Behrens, K. Bokeloh, E. Brown,J.M.R. Cardoso, B. Choi et al. , “Material screening and selection for XENON100,” *Astroparticle Physics,* vol. 35, p. 43–49, 2011.

[60] V. Alvarez, I. Bandac, A. Bettini, F. I. G. M. Borges, S. Cárcel, J. Castel, S. Cebrián, A. Cervera, C. A. N. Conde, T. Dafni et al. , “Radiopurity control in the NEXT-100 double beta decay experiment: procedures and initial measurements,” *Journal of Instrumentations,* vol. 8, p. T01002, 2013.

[61] L. Berge, R. S. Boiko, M. Chapellier, D. M. Chernyak, N. Coron, F. A. Danevich, R. Decourt, V. Ya Degoda, L. Devoyon, A. Drillien et al. , “Purification of molybdenum, growth and characterization of medium volume ZnMoO4 crystals for the LUMINEU program,” *Journal of Instrumentation,* vol. 9, p. P06004, 2014.

[62] B. Majorovits, H. Kader, H. Kraus, A. Lossin, E. Pantic, F. Petricca, F. Proebst, W. Seidel, “Production of low-background CuSn6-bronze for the CRESST dark-matter-

search experiment," *Applied Radiation and Isotopes,* vol. 67, pp. 197-200, 2009.

[63] F.A. Danevich, I. K. Bailiff, V. V. Kobychev, H. Kraus, M. Laubenstein, P. Loaiza, V. B. Mikhailik, S .S. Nagorny, A. S. Nikolaiko, S. Nisi et al., "Effect of recrystallisation on the radioactive contamination of CaWO4 crystal scintillators," *Nuclear Instruments and Methods in Physics Research A,* vol. 631, pp. 44-53, 2011.

[64] A. Munster, M. v. Sivers, G. Angloher, A. Bento, C. Bucci, L. Canonica, A. Erb, F. v. Feilitzsch, P. Gorla, A. Gutlein et al., "Radiopurity of CaWO4 crystals for direct dark matter search with CRESST and EURECA," *Journal of Cosmology and Astroparticle Physics* , vol. 5, p. 018, 2014.

[65] R. Strauss, G. Angloher, A. Bento, C. Bucci, L. Canonica, A. Erb, F.v. Feilitzsch, N. Ferreiro Iachellini, P. Gorla et al , "Beta/gamma and alpha backgrounds in CRESST-II Phase 2," *Journal of Cosmology and Astroparticle Physics* , vol. 06, p. 030, 2015.

[66] H. Kraus, M. Bauer, I. Bavykina, A. Benoit, J. Blümer, A. Broniatowski, V. Brudanin, G. Burghart, P. Camus, A. Chantelauze et al., "EURECA - the european future of Dark Matter searches with cryogenic detectors," *Nucl. Phys. B (Proc. Suppl.),* vol. 173, pp. 168-171, 2007.

[67] V. Bonvicini, S. Capelli, O. Cremonesi, G. Cucciati, L. Gironi, M. Pavan, E. Previtali, M. Sisti , "A flexible scintillation light apparatus for rare event searches," *The European Physical Journal C* , vol. 74, p. 3151, 2014.

[68] F.A. Danevich, A.Sh. Georgadze, V.V. Kobychev, B.N. Kropivyansky, S.S. Nagorny, A.S. Nikolaiko, D.V. Poda, V.I. Tretyak, I.M. Vyshnevskyi, S.S. Yurchenko et al., "Application of PbWO4 crystal scintillators in experiment to search for 2b decay of 116Cd," *Nuclear Instruments and Methods in Physics Research A* , vol. 556, p. 259–265, 2006.